\documentclass[twocolumn,showpacs,prl,amsmath,amssymb]{revtex4}
\usepackage{epsfig}
\usepackage{amsthm}
\usepackage{amssymb}
\usepackage{latexsym}
\usepackage{url}
\usepackage{amsmath, amscd}
\usepackage{verbatim}

\theoremstyle{definition}

\newcommand{\R}{\mathbb{R}}

\begin{document}

\title{On the motion of macroscopic bodies in quantum theory}
\author{A. \surname{Kryukov}} 
\affiliation{Department of Mathematics, University of Wisconsin Colleges, 34 Schroeder Ct, Madison, WI 53711} 
\begin{abstract}
Quantum observables can be identified with vector fields on the sphere of normalized states. The resulting {\it vector representation} is used in the paper to undertake a simultaneous treatment of macroscopic and microscopic bodies in quantum mechanics. Components of the velocity and acceleration of state under Schr{\"o}dinger evolution are given for a clear physical interpretation. Solutions to Schr{\"o}dinger and Newton equations are shown to be related beyond the Ehrenfest results on the motion of averages.  A formula relating the normal probability distribution and the Born rule is found. 
\end{abstract}


\maketitle

\section{Newtonian dynamics in Hilbert spaces}

Everyday experience shows that macroscopic bodies have well-defined position in space at any time. In the simplest case of a classical particle (material point) position at a given time is provided by vector ${\bf a}$ in the Euclidean space $\R^{3}$. Accordingly, the space $\R^{3}$ itself can be thought of as the space of all possible positions of a classical particle. 
In quantum mechanics the state of a spinless particle with a known position ${\bf a} \in \R^{3}$ is described by the Dirac delta function $\delta^{3}_{\bf a}({\bf x})=\delta^{3}({\bf x}-{\bf a})$. In particular, the state of a classical particle at any time is such a function.
The map $\omega: {\bf a} \longrightarrow \delta^{3}_{\bf a}$ provides a one-to-one correspondence between points ${\bf a} \in \R^{3}$ and ``state" functions $\delta^{3}_{\bf a}$. The set $\R^{3}$ can be then identified with the set $M_{3}$ of all delta functions in the space of state functions of the particle. 

The delta functions are, of course, not in the usual $L_{2}(\R^{3})$ Hilbert space on the measure $d^{3}x$. We study here a way to deal with functions of this type systematically and consistently, and in so doing, establish an interesting connection between the quantum theory and classical mechanics.

The inner product on the usual Hilbert space $L_{2}({\R}^{3})$ of state functions of a particle can be formally written for $\varphi, \psi \in L_{2}(\R^{3})$ in the following way:
\begin{equation}
\label{innerdd}
(\varphi, \psi)_{L_{2}}=\int \delta^{3}({\bf x}-{\bf y})\varphi({\bf x}){\overline \psi}({\bf y})d^{3}{\bf x}d^{3}{\bf y}. 
\end{equation}
In particular, the fact that delta functions are not in $L_{2}(\R^{3})$ is related to the singularity of delta functions. Let us replace the kernel $\delta^{3}({\bf x}-{\bf y})$  by the Gaussian function $\left(\frac{L}{\sqrt {2\pi}}\right)^{3}e^{-\frac{L^{2}}{2}({\bf x}-{\bf y})^{2}}$ for some positive constant $L$. This yields the product
\begin{equation}
\label{innerH}
(\varphi, \psi)_{\bf H}=\left(\frac{L}{\sqrt {2\pi}}\right)^{3}\int e^{-\frac{L^{2}}{2}({\bf x}-{\bf y})^{2}} \varphi({\bf x}){\overline \psi}({\bf y})d^{3}{\bf x}d^{3}{\bf y}.
\end{equation}
One can check Ref.\cite{KryukovIJMMS} that this is indeed an inner product on $L_{2}({\R}^{3})$. Some physical applications of this inner product were studied in Refs.\cite{KryukovJMP1}-\cite{KryukovFOP}. 
The separable Hilbert space ${\bf H}$ obtained by completing the space $L_{2}({\R}^{3})$ in this inner product contains delta functions  $\delta^{3}({\bf x}-{\bf a})$ and their derivatives. 
 Moreover, by choosing $L$ sufficiently large (or by choosing appropriate units), one can make the norm of any given square-integrable function in this metric as close as desired to its  $L_{2}({\R}^{3})$-norm. 

By dropping the coefficient $(1/{\sqrt {2\pi}})^{3}$ and using $L=\frac{1}{2\sigma}$ for an appropriate $\sigma$ we obtain the product 
\begin{equation}
\label{hh}
(\varphi, \psi)_{\bf H}=\int e^{-\frac{({\bf x}-{\bf y})^{2}}{8\sigma^{2}}}\varphi({\bf x})\overline{\psi}({\bf y})d^{3}{\bf x}d^{3}{\bf y}.
\end{equation}

Formally,
\begin{equation}
\label{delta1}
\int e^{-\frac{({\bf x}-{\bf y})^{2}}{8\sigma^{2}}}\delta^{3}({\bf x}-{\bf a})\delta^{3}({\bf y}-{\bf a})d^{3}{\bf x}d^{3}{\bf y}=1,
\end{equation}
so that the norm of the delta function  $\delta^{3}({\bf x}-{\bf a})$ in ${\bf H}$ with the metric (\ref{hh}) is $1$. 
The set $M_{3}$ of all delta functions  $\delta^{3}_{\bf a}({\bf x})$ with ${\bf a} \in \R^{3}$ is therefore a subset of the unit sphere in the Hilbert space ${\bf H}$. 

The map $\rho_{\sigma}: {\bf H} \longrightarrow L_{2}(\R^{3})$ that relates $L_{2}$ and ${\bf H}$-representations is given by the Gaussian kernel
\begin{equation}
\label{sigma}
\rho_{\sigma}({\bf x},{\bf y})=\left(\frac{1}{2\pi \sigma^{2}}\right)^{3/4}e^{-\frac{({\bf x}-{\bf y})^{2}}{4\sigma^{2}}}.
\end{equation}
In terms of $\rho_{\sigma}$, the kernel $G({\bf x}, {\bf y})$ of the metric on ${\bf H}$ is given by
\begin{equation}
\label{GG}
G({\bf x}, {\bf y})=(\rho^{\ast}_{\sigma}\rho_{\sigma})({\bf x}, {\bf y})=e^{-\frac{({\bf x}-{\bf y})^{2}}{8\sigma^{2}}},
\end{equation}
which is consistent with (\ref{hh}).
The map $\rho_{\sigma}$ transforms delta functions $\delta^{3}_{\bf a}$ to Gaussian functions ${\widetilde \delta^{3}_{\bf a}}=\rho_{\sigma}(\delta^{3}_{\bf a})$, centered at ${\bf a}$, providing an alternative, more common way of dealing with singularity of delta functions. The image $M^{\sigma}_{3}$ of $M_{3}$ under $\rho_{\sigma}$ is a subset of the unit sphere in $L_{2}(\R^{3})$ made of the functions ${\widetilde \delta^{3}_{\bf a}}$. Both realizations will prove useful in the discussion of motion of macroscopic bodies in quantum mechanics.

To know position ${\bf a}$ of a classical particle in $\R^{3}$ is to know the corresponding point $\delta^{3}_{\bf a}$ in $M_{3}$. 
Consider a path ${\bf r}={\bf a}(t)$ with values in $\R^{3}$ and the corresponding path $\varphi=\delta^{3}_{{\bf a}(t)}$ in $M_{3}$.
With the use of the chain rule the velocity vector $d \varphi/dt$ can be written as
\begin{equation}
\label{chain}
\frac{d \varphi}{dt}=-\frac{\partial}{\partial {\bf x}^{i}}\delta^{3}({\bf x}-{\bf a})\frac {d{\bf a}^{i}}{dt},
\end{equation}
where the usual summation convention for repeating indices is accepted.
It follows that the norm $\left \| \frac{d \varphi}{dt} \right \|^{2}_{H}$ of the velocity in the space ${\bf H}$ is
\begin{equation}
\int k({\bf x},{\bf y}) \frac{\partial}{\partial x^{i}}\delta^{3}({\bf x}-{\bf a})\frac {d{\bf a}^{i}}{dt}\frac{\partial}{\partial y^{k}}\delta^{3}({\bf y}-{\bf a})
\frac {d{\bf a}^{k}}{dt}d^{3}{\bf x}d^{3}{\bf y},
\end{equation}
where $k({\bf x},{\bf y})=e^{-\frac{({\bf x}-{\bf y})^{2}}{8\sigma^{2}}}$.
``Integration by parts" in the last expression gives
\begin{equation}
\label{parts1}
\left \| \frac{d \varphi}{dt} \right \|^{2}_{H}=
\left.\frac {\partial^{2}k({\bf x},{\bf y})}{\partial x^{i} \partial y^{k}}\right|_{{\bf x}={\bf y}={\bf a}} \frac {d{\bf a}^{i}}{dt}\frac {d{\bf a}^{k}}{dt}.
\end{equation}
Furthermore, 
\begin{equation}
\left.\frac {\partial^{2}k({\bf x},{\bf y})}{\partial x^{i} \partial y^{k}}\right|_{{\bf x}={\bf y}={\bf a}}=\frac{1}{4\sigma^{2}}\delta_{ik},
\end{equation}
where $\delta_{ik}$ is the Kronecker delta symbol. Assuming now that the distance in $\R^{3}$ is measured in the units of $2\sigma$ (equivalently, taking $2\sigma=1$) one obtains the equality of the speeds
\begin{equation}
\label{Norms}
\left \| \frac{d \varphi}{dt} \right \|_{H}=\left \| \frac{d {\bf a}}{dt} \right \|_{\R^{3}}.
\end{equation}
From this equality of norms it follows that the set $M_{3}$ as a metric subspace of ${\bf H}$ is identical to the Euclidean space $\R^{3}$. That is, the one-to-one map $\omega: \R^{3} \longrightarrow {\bf H}$ is an isometric embedding Ref.\cite{KryukovIJMMS}. Notice however that $M_{3}$ is not a vector subspace of ${\bf H}$. Rather, as follows from (\ref{delta1}), the metric space $M_{3}$ is a submanifold of the unit sphere $S^{\bf H}$ in ${\bf H}$. Since delta functions $\delta^{3}_{{\bf a}_{k}}$ with different ${{\bf a}_{k}}$, $k=1,...,n$ are linearly independent, the manifold $M_{3}$ ``spirals'' through dimensions of the sphere, forming a complete subset of ${\bf H}$. This means that no function in ${\bf H}$ is orthogonal to the submanifold $M_{3}$ Ref.\cite{KryukovIJMMS}. 

Nevertheless, a vector structure on $M_{3}$ exists. For instance, define the operations of addition $\oplus$ and multiplication by a scalar $\lambda \odot$  via $\omega({\bf a})\oplus\omega({\bf b})=\omega({\bf a}+{\bf b})$ and $\lambda \odot\omega({\bf a})=\omega(\lambda {\bf a})$, where the map $\omega$ is the same as before. The resulting operations are continuous in the topology of $M_{3}\subset {\bf H}$. That is, the metric space $M_{3}$ with this vector structure is isomorphic to the vector space $\R^{3}$ with the Euclidean metric. 

From
\begin{equation}
\frac{d}{dt}\delta^{3}_{{\bf a}}({\bf x})=-\frac{\partial}{\partial x^{i}} \delta^{3}_{{\bf a}}({\bf x})\frac{d a^{i}}{dt}
\end{equation}
and 
\begin{equation}
\frac{d^{2}}{dt^{2}}\delta^{3}_{{\bf a}}({\bf x}) =
\frac{\partial^{2}}{\partial x^{i}\partial x^{j}} \delta^{3}_{{\bf a}}({\bf x})\frac{d a^{i}}{dt}\frac{d a^{j}}{dt}-\frac{\partial}{\partial x^{i}} \delta^{3}_{{\bf a}}({\bf x})\frac{d^{2} a^{i}}{dt^{2}},
\end{equation}
together with (\ref{delta1}), (\ref{Norms}), and the orthogonality of the first and second derivatives of $\delta^{3}_{\bf a}({\bf x})$, it follows that projection of velocity and acceleration of the state $\delta^{3}_{{\bf a}(t)}$ onto $M_{3}$ yields correct Newtonian velocity and acceleration of the classical particle. That is:
\begin{equation}
\label{v1}
\left( \frac{d}{dt}\delta^{3}_{{\bf a}}({\bf x}), -\frac{\partial}{\partial x^{i}} \delta^{3}_{{\bf a}}({\bf x})\right)_{\bf H} =\frac{d a^{i}}{dt}
\end{equation}
and 
\begin{equation}
\label{a1}
\left( \frac{d^{2}}{dt^{2}}\delta^{3}_{{\bf a}}({\bf x}) , -\frac{\partial}{\partial x^{i}} \delta^{3}_{{\bf a}}({\bf x})\right)_{\bf H} =\frac{d^{2} a^{i}}{dt^{2}}.
\end{equation}

Furthermore, Newtonian dynamics of the classical particle follows from the principle of least action for the action functional $S$ on paths in ${\bf H}$, defined by
\begin{equation}
\int k({\bf x},{\bf y})\left[\frac{m}{2}\frac{d \varphi_{t}({\bf x})}{dt} \frac{d{\overline  \varphi_{t}}({\bf y})}{dt}-V({\bf x}) \varphi_{t}({\bf x}) {\overline \varphi_{t}}({\bf y})\right]d^{3}{\bf x}d^{3}{\bf y}dt,
\end{equation}
where $m$ is the mass of the particle, $V$ is the potential and $k({\bf x}, {\bf y})=e^{-\frac{1}{2}({\bf x}-{\bf y})^{2}}$, as before. Suppose that $\varphi_{t}$ is constrained to take values on the submanifold $M_{3}\subset {\bf H}$, i.e., $\varphi_{t}({\bf x})=\delta^{3}({\bf x}-{\bf a}(t))$. Using (\ref{chain}) and integrating by parts as in (\ref{parts1}), we immediately obtain
\begin{equation}
S=\int\left[\frac{m}{2}\left(\frac{d{\bf a}}{dt}\right)^{2}-V({\bf a})\right]dt,
\end{equation}
i.e., the usual action functional for a material point in classical mechanics. In these terms, a classical particle is a constrained dynamical system in ${\bf H}$. The same applies to $L_{2}(\R^{3})$-representation and paths constrained to take values in $M^{\sigma}_{3}=\rho_{\sigma}(M_{3})$ in $L_{2}(\R^{3})$.

Classical particle mechanics, therefore, has an equivalent realization in terms of the new dynamical variables: the state $\varphi$ of the particle and the velocity $\frac{d\varphi}{dt}$ of the state. A similar realization exists for mechanical systems consisting of any number of classical particles. For example, the map $\omega \otimes \omega: \R^{3}\times \R^{3} \longrightarrow {\bf H}\otimes {\bf H}$, $\omega \otimes \omega ({\bf a}, {\bf b})=\delta^{3}_{\bf a} \otimes \delta^{3}_{\bf b}$ identifies the configuration space $\R^{3}\times \R^{3}$ of a two particle system with the embedded submanifold $M_{6}=\omega \otimes \omega(\R^{3}\times \R^{3})$ of the Hilbert space ${\bf H}\otimes {\bf H}$. Consider a path $({\bf a}(t), {\bf b}(t))$ in $\R^{3}\times \R^{3}$ and the corresponding path $\delta^{3}_{{\bf a}(t)}\otimes \delta^{3}_{{\bf b}(t)}$ with values in $M_{6}$. For any $t$, the vectors $\frac{d}{dt}\delta^{3}_{{\bf a}(t)}\otimes \delta^{3}_{{\bf b}(t)}$ and $\delta^{3}_{{\bf a}(t)}\otimes \frac{d}{dt}\delta^{3}_{{\bf b}(t)}$ are tangent to $M_{6}$ at the point $\delta^{3}_{{\bf a}(t)}\otimes \delta^{3}_{{\bf b}(t)}$ and orthogonal in ${\bf H}\otimes {\bf H}$. The space $M_{6}$ with the induced metric is isometric to the direct product $\R^{3}\times \R^{3}$ with the natural Euclidean metric. Projection of velocity and acceleration of the state $\varphi(t)=\delta^{3}_{{\bf a}(t)}\otimes \delta^{3}_{{\bf b}(t)}$ onto the basis vectors $\left(- \frac{\partial}{\partial x^i}\delta^{3}_{{\bf a}(t)}\right)\otimes \delta^{3}_{{\bf b}(t)}$ and $\delta^{3}_{{\bf a}(t)}\otimes \left(-\frac{\partial}{\partial x^{k}}\delta^{3}_{{\bf b}(t)}\right)$ yields the velocity and acceleration of the particles by means of the formulas similar to (\ref{v1}) and (\ref{a1}).

We now turn the attention to quantum theory and explore a useful realization of quantum mechanics in terms of vector fields in the space of states.

\section{Observables as vector fields}


Quantum observables can be identified with vector fields on the space of states Ref.\cite{KryukovUncert}. Namely, given a self-adjoint operator ${\widehat A}$ on a Hilbert space $L_{2}$ of square-integrable functions (it could in particular be the tensor product space of a many body problem)  one can introduce the associated linear vector field $A_{\varphi}$ on $L_{2}$ by
\begin{equation}
\label{vector}
A_{\varphi}=-i{\widehat A}\varphi.
\end{equation}
This field is defined on a dense subset $D$ in $L_{2}$ on which the operator ${\widehat A}$ itself is defined.  Clearly, to know the vector field $A_{\varphi}$ is the same as to know the operator ${\widehat A}$ itself.
Moreover, the commutator of observables and the commutator (Lie bracket) of the corresponding vector fields are related in a simple way:
\begin{equation}
\label{comm}
[A_{\varphi},B_{\varphi}]=[{\widehat A},{\widehat B}]\varphi.
\end{equation}

The field $A_{\varphi}$ associated with an observable, being restricted to the sphere $S^{L_{2}}$ of unit normalized states, is tangent to the sphere.
Indeed, the equation for the integral curves of $A_{\varphi}$ has the form
\begin{equation}
\label{SchroedA}
\frac{d \varphi_{\tau}}{d\tau}=-i{\widehat A}\varphi_{\tau}.
\end{equation}
The solution to (\ref{SchroedA}) through initial point $\varphi_{0}$ is given by 
$\varphi_{\tau}=e^{-i{\widehat A}\tau}\varphi_{0}$.
Here $e^{-i{\widehat A}\tau}$ denotes the one-parameter group of unitary transformations generated by $-i{\widehat A}$, as described by Stone's theorem.
It follows that the integral curve through $\varphi_{0} \in S^{L_{2}}$ will stay on the sphere. One concludes that, modulo the domain issues, the restriction of the vector field $A_{\varphi}$ to the sphere $S^{L_{2}}$ is a vector field on the sphere.

Under the embedding, the inner product on the Hilbert space $L_{2}$ gives rise to a Riemannian metric (i.e., point-dependent real-valued inner product) on the sphere $S^{L_{2}}$. For this one considers the realization $L_{2R}$ of the Hilbert space $L_{2}$, i.e., the real vector space of pairs $X=({\mathrm Re} \psi, {\mathrm Im} \psi)$ with $\psi$ in $L_{2}$. If $\xi, \eta$ are vector fields on $S^{L_{2}}$, 
one can define a Riemannian metric $G_{\varphi}: T_{R\varphi}S^{L_{2}}\times T_{R\varphi}S^{L_{2}} \longrightarrow R$ on the sphere by
\begin{equation}
\label{Riem}
G_{\varphi}(X,Y)={\mathrm Re} (\xi, \eta).
\end{equation}
Here the tangent space $T_{R\varphi}S^{L_{2}}$ to $S^{L_{2}}$ at a point $\varphi$ is identified with an affine subspace in $L_{2R}$, $X=({\mathrm Re} \xi, {\mathrm Im} \xi)$, $Y=({\mathrm Re} \eta, {\mathrm Im} \eta)$ and $(\xi, \eta)$ denotes the $L_{2}$-inner product of $\xi, \eta$. 
Note that the obtained Riemannian metric $G_{\varphi}$ is {\em strong} in the sense that it yields an isomorphism ${\widehat G}:T_{R\varphi}S^{L_{2}}\longrightarrow \left (T_{R\varphi}S^{L_{2}}\right)^{\ast}$ of dual spaces.   

The Riemannian metric on $S^{L_{2}}$ yields a (strong) Riemannian metric on the projective space $CP^{L_{2}}$. For this, one defines the metric on $CP^{L_{2}}$ so that the bundle projection $\pi: S^{L_{2}} \longrightarrow CP^{L_{2}}$ would be a Riemannian submersion. The resulting metric on $CP^{L_{2}}$ is called the Fubini-Study metric. To put it simply, 
an arbitrary tangent vector $X \in T_{R\varphi}S^{L_{2}}$
can be decomposed into two components: tangent and orthogonal to the fibre $\{\varphi\}$ through $\varphi$ (i.e., to the plane $C^{1}$ containing the circle $S^{1}=\{\varphi\}$). The differential $d\pi$ maps the tangent component to the zero-vector. The orthogonal component of $X$ can be then identified with $d\pi(X)$. 
If two vectors $X,Y$ are orthogonal to the fibre $\{\varphi\}$, the inner product of $d\pi(X)$ and $d\pi(Y)$ in the Fubini-Study metric is equal to the inner product of $X$ and $Y$ in the metric $G_{\varphi}$. Note that the obtained Riemannian metrics on $S^{L_{2}}$ and $CP^{L_{2}}$ are invariant under the induced action of the group of unitary transformations on $L_{2}$.

An arbitrary vector in the Hilbert space at a point $\varphi$ can be decomposed onto the radial component (parallel to the radius vector from the origin to the point $\varphi$, i.e., parallel to $\varphi$ itself), and tangential component. 
The radial component of a vector field $A_{\varphi}$ associated with an observable vanishes. Accordingly, $A_{\varphi}$ can be decomposed into components tangent and orthogonal to the fibre $\{\varphi\}$. These components have a simple physical meaning.
In fact, the equality
\begin{equation}
{\overline A} \equiv (\varphi, {\widehat A}\varphi)=(-i\varphi, -i{\widehat A}\varphi),
\end{equation}
signifies that the expected value of an observable ${\widehat A}$ in the state $\varphi$ is the projection of the vector $-i{\widehat A}\varphi \in T_{\varphi}S^{L_{2}}$ on the unit vector $-i\varphi=-i I \varphi \in T_{\varphi}S^{L_{2}}$, tangent to the fibre $\{\varphi\}$.
Because
\begin{equation}
(\varphi, {\widehat A}^{2}\varphi)=({\widehat A}\varphi, {\widehat A}\varphi)=(-i{\widehat A}\varphi, -i{\widehat A}\varphi),
\end{equation}
the term $(\varphi, {\widehat A}^{2}\varphi)$ is just the norm of the vector $-i{\widehat A}\varphi$ squared. The expected value $(\varphi, {\widehat A}_{\bot}\varphi)$ of the operator ${\widehat A}_{\bot} \equiv {\widehat A}-{\overline A}I$ in the state $\varphi$ is zero. Therefore, the vector $-i{\widehat A}_{\bot}\varphi=-i{\widehat A}\varphi-(-i{\overline A}\varphi)$, which is the component of $-i{\widehat A}\varphi$ orthogonal to $-i\varphi$  is  orthogonal to the fibre $\{\varphi\}$.
Accordingly, the variance 
\begin{equation}
\Delta A^{2}=(\varphi, ({\widehat A}-{\overline A}I)^{2}\varphi)=(\varphi, {\widehat A}_{\bot}^{2}\varphi)=(-i{\widehat A}_{\bot}\varphi, -i{\widehat A}_{\bot}\varphi) 
\end{equation}
is the norm squared of the component $-i{\widehat A}_{\bot}\varphi$. As discussed, the image of this vector under $d\pi$ can be identified with the vector itself. 
It follows that the norm of $-i{\widehat A}_{\bot}\varphi$ in the Fubini-Study metric coincides with its norm in the Riemannian metric on $S^{L_{2}}$ (and in the original $L_{2}$-metric).

Integral curves of the vector field $A_{\varphi}=-i{\widehat A} \varphi$ are solutions to the equation
\begin{equation}
\label{evoll}
\frac{d\varphi}{dt}=-i{\widehat A}\varphi
\end{equation}
for the state $\varphi$ with the initial condition $\left.\varphi \right|_{t=0}=\varphi_{0}$.

Decomposition of $-i{\widehat A}\varphi$ onto the components parallel   and orthogonal to the fibre yields the equation 
\begin{equation}
\label{evolll}
\frac{d\varphi}{dt}=-i{\overline A}\varphi+\left(-i{\widehat A}\varphi+i{\overline A}\varphi\right)=
-i{\overline A}\varphi-i {\widehat A}_{\perp}\varphi.
\end{equation}
By projecting both sides of this equation by $d\pi$ one obtains
\begin{equation}
\label{ddt}
\frac{d\{\varphi\}}{dt}=-i{\widehat A}_{\bot}\varphi.
\end{equation}

The left hand side of (\ref{ddt}) is the velocity of evolution of the projection $\{\varphi\}=\pi(\varphi)$ in  $CP^{L_{2}}$. By the above, the norm of the right hand side is the uncertainty of ${\widehat A}$ in the state $\varphi$:
\begin{equation}
\label{speed}
\|-i{\widehat A}_{\bot}\varphi\|=\Delta A.
\end{equation}

In particular, if ${\widehat A}$ is the Hamiltonian ${\widehat h}$, then equation (\ref{evoll}) is the Scr{\"o}dinger equation and the following result is obtained:
{\it The velocity of evolution of state in the projective space is equal to the uncertainty of energy.} 
This result was obtained first in Ref.\cite{AA} by using different methods.

Now let's decompose the acceleration vector $\frac{d^{2}\varphi}{dt^{2}}=\frac{d}{dt}\left( -i{\widehat h}\varphi\right)=-{\widehat h}^{2}\varphi$. Notice first of all that
\begin{equation}
{\mathrm Re}(-i\varphi, {\widehat h}^{2}\varphi)=0,
\end{equation}
so that the parallel tangential component of acceleration of Shr{\"o}dinger evolution vanishes. This simply means that the phase component of the velocity (i.e., the expected value of energy, see above) does not change. In particular, the tangential component is purely orthogonal. The radial component is given by $-(\varphi, {\widehat h}^{2}\varphi)\varphi=-(-i{\widehat h}\varphi, -i{\widehat h}\varphi)\varphi$. Since $-i{\widehat h}\varphi$ is the velocity of evolution, we recognize in this term the centropidical acceleration ($-\frac{{\bf v}^{2}{\bf r}}{r^{2}}$ with $r=1$).

The tangential component is therefore equal to 
\begin{equation}
-{\widehat h}^{2}\varphi + (\varphi,{\widehat h}^{2}\varphi)\varphi=-{\widehat h}^{2}_{\perp}\varphi.
\end{equation}
Therefore, the following result is obtained: {\it Acceleration of the Schro{\"o}dinger evolution of state in the projective space is equal to the uncertainty of the square of energy.}


\section{Components of velocity of state}


Classical and quantum mechanics of a particle are now formulated within the same Hilbert space framework. Recall that the space $\R^{3}$ is now identified via the map $\omega$ with the submanifold $M_{3}$ in ${\bf H}$ with the induced Euclidean metric. Alternatively, the map $\omega_{\sigma}=\rho_{\sigma}\omega$ identifies $\R^{3}$ with the submanifold $M^{\sigma}_{3}$ in $L_{2}(\R^{3})$. This later equivalent realization will be used in this section.
Note that because all normalized Gaussian functions of a given width $\sigma$ are obtained from a single one by translations in ${\bf x}$, the field ${\bf p}_{\varphi}=-i{\widehat {\bf p}}\varphi$ for $\varphi \in M^{\sigma}_{3}$ is tangent to $M^{\sigma}_{3}$. 
The goal here is to use the embedding $\omega_{\sigma}$ of $\R^{3}$ into the space of states together with the vector representation of observables to study  the relation of the Schr{\"o}dinger evolution with the classical Newtonian motion.



One standard way to describe this relation is via the Ehrenfest theorem (the expected value of the Heisenberg equation of motion):
\begin{equation}
\label{Her}
\frac{d}{dt}(\varphi, {\widehat A}\varphi)=-i(\varphi, [{\widehat A}, {\widehat h}]\varphi).
\end{equation}
Here ${\widehat A}$ does not depend on $t$. For example, for the momentum operator of a free particle we obtain
\begin{equation}
\label{para}
\frac{d{\overline{\bf p}}}{dt}=0.
\end{equation}
Recall that ${\overline{\bf p}}$ is the phase projection of the vector field $p_{\varphi}$. The equation (\ref{para}) simply says that this projection is time-independent. Note that the orthogonal projection, i.e. the uncertainty $\Delta{\bf p}$ is also preserved in this case and this is not captured in (\ref{Her}).

Compare (\ref{Her})  to another equation that follows from the Schr{\"o}dinger dynamics:
\begin{equation}
\label{projj}
 2\left(\frac{d  \varphi}{dt}, -i {\widehat A} \varphi \right)=
 \left( \varphi, \{{\widehat A}, {\widehat h} \}\varphi \right)-\left(\varphi,[{\widehat A}, {\widehat h}]\varphi\right).
\end{equation}
The Ehrenfest theorem (\ref{Her}) for a time-independent observable amounts to using  the imaginary part of (\ref{projj}), i.e., the part with the commutator $[{\widehat A}, {\widehat h}]$. The left hand side of (\ref{projj}) is twice the projection of the velocity of state onto the vector field associated with the observable ${\widehat A}$. The real part of this projection (the term with the anticommutator $\{{\widehat A}, {\widehat h}\}$) is twice the projection in the sense of Riemannian metric (\ref{Riem}). This Riemannian projection will be used here.

Suppose that at $t=0$ a microscopic particle is prepared in the state
\begin{equation}
\label{initial}
    \varphi_{0}({\bf x})=\left(\frac{1}{2\pi\sigma^{2}}\right)^{3/4}e^{-\frac{({\bf x}-{\bf x}_{0})^{2}}{4\sigma^{2}}}e^{i\frac{{\bf p}_{0}({\bf x}-{\bf x}_{0})}{\hbar}},
\end{equation}
where $\sigma$ is the same as in (\ref{sigma}) and ${\bf p}_{0}=m{\bf v}_{0}$ with ${\bf v}_{0}$ being the initial group-velocity of the packet. 
The set of all initial states $\varphi_{0}$ given by (\ref{initial}) form a $6$-dimensional embedded submanifold $M^{\sigma}_{3,3}$ in $L_{2}(\R^{3})$. 
The map $\Omega: \R^{3}\times \R^{3} \longrightarrow M^{\sigma}_{3,3}$,
\begin{equation}
\Omega({\bf a},{\bf p})=\left(\frac{1}{2\pi\sigma^{2}}\right)^{3/4}e^{-\frac{({\bf x}-{\bf a})^{2}}{4\sigma^{2}}}e^{i\frac{{\bf p}({\bf x}-{\bf a})}{\hbar}}
\end{equation} is a diffeomorphism from the classical phase space of the particle onto the manifold $M^{\sigma}_{3,3}$. 
For any path $\varphi=\varphi_{\tau}$ in $L_{2}(\R^{3})$, $\varphi=re^{i\theta}$, the terms of the derivative
\begin{equation}
\frac{d\varphi}{d\tau}=\frac{d r}{d\tau}e^{i\theta}+i\frac{d \theta}{d\tau}re^{i\theta}
\end{equation}
are orthogonal in the Riemannian metric:
\begin{equation}
\label{orthog}
\mathrm{Re}\left(\frac{dr}{d\tau}e^{i\theta}, i\frac{d\theta}{d\tau}re^{i\theta}\right)=0.
\end{equation}
In particular, the vectors $\frac{\partial r}{\partial x^{\alpha}}e^{i\theta}$ and $i\frac{\partial \theta}{\partial p^{\beta}}re^{i\theta}$ tangent to the manifold $M^{\sigma}_{3,3}$ at a point $\varphi_{0}$ are orthogonal and form a basis in the tangent space at that point. For any path $\varphi_{\tau}$ with values in $M^{\sigma}_{3,3}$ the norm of velocity vector $\frac{d \varphi}{d\tau}$ is given by
\begin{equation}
\label{phaseMetric}
\left\|\frac{d \varphi}{d \tau}\right\|^{2}_{L_{2}}=\frac{1}{4\sigma^{2}}\left\|\frac{d {\bf a}}{d\tau}\right\|^{2}_{\R^{3}}+\frac{\sigma^{2}}{\hbar^{2}}\left\|\frac{d {\bf p}}{d\tau}\right\|^{2}_{\R^{3}}.
\end{equation}
That is, under a proper choice of units, the map $\Omega$ is an isometry, which identifies the Euclidean phase space $\R^{3}\times \R^{3}$ of the particle with the embedded submanifold $M^{\sigma}_{3,3} \subset L_{2}(\R^{3})$ furnished with the induced Riemannian metric. The map $\Omega$ is an extension to the phase space of the isometric embedding $\omega_{\sigma}=\rho_{\sigma}\circ\omega$ of the space $\R^{3}$ considered in the first section. 

Suppose that the state (\ref{initial}) evolves according to the Schr{\"o}dinger equation with the Hamiltonian ${\widehat h}=-\frac{\hbar^{2}}{2m}\Delta+V({\bf x})$. 
At any point $\varphi_{0} \in M^{\sigma}_{3,3}$, the velocity vector $\frac{d\varphi}{dt}$ is tangent to the unit sphere of states $S^{L_{2}}$ in $L_{2}(\R^{3})$ and can be decomposed into a sum of components of physical interest. First of all, by (\ref{evolll}) 
\begin{equation}
\frac{d\varphi}{dt}=-\frac{i}{\hbar}{\widehat h}\varphi=-\frac{i}{\hbar}{\overline E}\varphi-\frac{i}{\hbar}{\widehat h}_{\perp}\varphi.
\end{equation}
So, once again,  the component of $\frac{d\varphi}{dt}$ along the vector $i\varphi$ is $\frac{\overline E}{\hbar}$ and the norm of the orthogonal component $\left\|-\frac{i}{\hbar}{\widehat h}_{\perp}\varphi\right\|$ is  $\frac{\Delta h}{\hbar}$.

To decompose the orthogonal component $-\frac{i}{\hbar}{\widehat h}_{\perp}\varphi$ of the velocity $\frac{d\varphi}{dt}$, notice that 
the orthogonal vectors $\frac{\partial r}{\partial x^{\alpha}}e^{i\theta}$ and $i\frac{\partial \theta}{\partial p^{\beta}}re^{i\theta}$ tangent to $M^{\sigma}_{3,3}$ are also orthogonal to vector $i\varphi$:
\begin{equation}
{\mathrm Re}\left(i\varphi, -\frac{\partial r}{\partial x^{\alpha}}e^{i\theta}\right)=0 \ \textnormal{for all} \ t
\end{equation}
and 
\begin{equation}
\left(i\varphi, i\frac{\partial\theta}{\partial p^{\alpha}}\varphi\right)=0 \  \textnormal{for} \ t=0.
\end{equation}
Calculation of the projection of the velocity $\frac{d \varphi}{dt}$ onto the unit vector $-\widehat{\frac{\partial r}{\partial x^{\alpha}}}e^{i\theta}$ (i.e., the classical space component of $\frac{d\varphi}{dt}$) for any Hamiltonian ${\widehat h}=-\frac{\hbar^{2}}{2m}\Delta+V({\bf x})$ yields
\begin{equation}
\label{pproj}
\left.\mathrm{Re}\left(\frac{d \varphi}{dt}, -\widehat{ \frac{\partial r}{\partial x^{\alpha}}}e^{i\theta}\right)\right|_{t=0}=\left.\left(\frac{d r}{dt}, -\widehat{ \frac{\partial r}{\partial x^{\alpha}}}\right)\right|_{t=0}=\frac{v^{\alpha}_{0}}{2\sigma}.
\end{equation}
Calculation of the projection of velocity $\frac{d \varphi}{dt}$ onto the unit vector $i\widehat{\frac{\partial\theta}{\partial p^{\alpha}}}\varphi$ (momentum space component) gives
\begin{equation}
\label{w}
\left.\mathrm{Re} \left(\frac{d\varphi}{dt}, i\widehat{\frac{\partial\theta}{\partial p^{\alpha}}}\varphi\right)\right|_{t=0}=\frac{mw^{\alpha} \sigma}{\hbar},
\end{equation}
where
\begin{equation}
mw^{\alpha}=-\left.\frac{\partial V({\bf x})}{\partial x^{\alpha}}\right|_{{\bf x}={\bf x}_{0}}
\end{equation}
and $\sigma$ is assumed to be small enough for the linear approximation for $V({\bf x})$ to be valid within intervals of length $\sigma$. 

The velocity $\frac{d\varphi}{dt}$ also contains component which is due to the change in $\sigma$ (spreading). The inner product
\begin{equation}
\left(i\varphi, i\frac{d\varphi}{d\sigma}\right)=\left(\varphi, \frac{d\varphi}{d\sigma}\right)
\end{equation}
vanishes at $t=0$, so the vector $i\frac{d\varphi}{d\sigma}$ is also tangent to the sphere $S^{L_{2}}$ and orthogonal to the phase circle. It is also orthogonal to the phase space $M^{\sigma}_{3,3}$. The component of the velocity $\frac{d\varphi}{dt}$ along this vector is given by
\begin{equation}
\label{spreadcomp}
\left.\mathrm{Re} \left (\frac{d\varphi}{dt}, i\widehat{\frac{d\varphi}{d\sigma}}\right)\right|_{t=0}=\frac{\sqrt{2}\hbar}{8\sigma^{2}m}.
\end{equation}
Finally, calculation of the norm of $\frac{d\varphi}{dt}=\frac{i}{\hbar}{\widehat h}\varphi$ at $t=0$ gives
\begin{equation}
\label{decomposition}
\left\|\frac{d\varphi}{dt}\right\|^{2}=\frac{{\overline E}^{2}}{\hbar^{2}}+\frac{{\bf v}^{2}_{0}}{4\sigma^{2}}+\frac{m^{2}{\bf w}^{2}{\sigma}^{2}}{\hbar^{2}}+\frac{\hbar^{2}}{32\sigma^{4}m^{2}},
\end{equation}
which is exactly the sum of squares of the found components. This, therefore, completes a decomposition of the velocity of state at any point $\varphi_{0} \in M^{\sigma}_{3,3}$.

Note that for a closed system the norm  of $\frac{d\varphi}{dt}=\frac{i}{\hbar}{\widehat h}\varphi$ is preserved in time. For a system in a stationary state, this amounts to conservation of energy. In fact, in this case $\varphi_{t}({\bf x})=\psi({\bf x})e^{-\frac{iEt}{\hbar}}$, which is a motion along the phase circle, and
\begin{equation}
\left\| \frac{d\varphi}{dt}\right\|^{2}=\frac{E^2}{\hbar^2}.
\end{equation}
As discussed in the section titled Observables as vector fields, for any initial state the norm of the phase component (expected energy) and orthogonal component (energy uncertainty) of the velocity $\frac{d\varphi}{dt}$ are both preserved. 

The presence of $i\frac{d\varphi}{d\sigma}$ component of the the velocity in (\ref{decomposition}) hints that the classical phase space $M^{\sigma}_{3,3}$ may be usefully extended to include all positive values of $\sigma$. The induced metric on the resulting manifold $M^{\sigma}_{3,3} \times \R_{+}$ is then given by the following extension of (\ref{phaseMetric}):
\begin{equation}
\label{phaseMetric1}
\left\|\frac{d \varphi}{dt}\right\|^{2}_{L_{2}}=\frac{1}{4\sigma^{2}}\left\|\frac{d {\bf a}}{dt}\right\|^{2}_{\R^{3}}+\frac{\sigma^{2}}{\hbar^{2}}\left\|\frac{d {\bf p}}{dt}\right\|^{2}_{\R^{3}}+\frac{1}{2\sigma^{2}}\left|\frac{d\sigma}{dt}\right|^{2}.
\end{equation}
With appropriate units, this gives an isometric embedding of the extended phase space $\R^{3} \times \R^{3} \times \R_{+}$ with Euclidean metric into $L_{2}(\R^{3})$.

The "spreading" component of the velocity admits an interesting interpretation. Suppose that the width of the initial state $\varphi_{0}$ is given by the Compton length $\frac{\hbar}{mc}$, which is a natural limit on the width of state in quantum mechanics. From this and (\ref{spreadcomp}) and (\ref{decomposition}) it follows that component of velocity of state due to spreading is proportional to the mass $m$ of the particle. So the mass can be thought of as the speed of motion of state in the direction of spreading, orthogonal to the phase space $M^{\sigma}_{3,3}$. The sum of the last three terms in (\ref{decomposition}) is equal to the square of the uncertainty $\Delta h$. If ${\bf v}_{0}$ and ${\bf w}$ vanish, then 
\begin{equation}
\Delta h \ = \ \textnormal{mass term} \ = \ \textnormal{speed of spreading}.
\end{equation}

In the linear potential approximation, the first term in (\ref{decomposition}) is the square of the term
\begin{equation}
\label{restmass}
\frac{1}{\hbar}\left(U+K+\frac{\hbar^{2}}{8m\sigma^{2}}\right),
\end{equation}
where $U=V({\bf x}_{g})$ and $K=\frac{m{\bf v}^{2}_{g}}{2}$ are potential and kinetic energy of the packet considered as a particle with position ${\bf x}_{g}={\bf x}_{0}+{\bf v}_{0}t+\frac{{\bf w}t^{2}}{2}$ and velocity ${\bf v}_{g}={\bf v}_{0}+{\bf w}t$. The last term in parentheses in (\ref{restmass}) accounts for the difference in energy of the packets with the same $U$ and $K$, but different values of $\sigma$. Up to a constant factor this term equals the component of velocity due to spreading given by (\ref{spreadcomp}). With the unit of length $2\sigma$ given by Compton length and the choice of units that make the metric (\ref{phaseMetric1}) for a particle of a given mass Euclidean, this term is equal to the rest energy $mc^{2}$ of the particle. 

Calculations show that for $t>0$ the spatial component (\ref{pproj}) of velocity of state is given by  $\frac{v^{\alpha}_{g}}{2\sigma_{t}}$ while the component (\ref{spreadcomp}) due to spreading does not change. Here ${\bf v}_{g}={\bf v}_{0}+{\bf w}t$ is the group velocity and $\sigma_{t}$, given by
\begin{equation}
\label{sigmaS}
\sigma^{2}_{t}=\sigma^{2}\left(1+\frac{\hbar^{2}t^{2}}{4m^{2}\sigma^{4}}\right),
\end{equation}
is the width of the packet at time $t$, and it is assumed that $\sigma_{t}$ is sufficiently small for the linear approximation of $V({\bf x})$ to be valid. 
The relationship
\begin{equation}
\label{Va}
\frac{d}{dt}\left.\left(\frac{d r}{dt}, - \widehat{\frac{\partial r}{\partial{x}^{\alpha}}} \right)\right |_{t=0} 
=-\left.\frac{1}{m}\frac{\partial V({\bf x})}{\partial x^{\alpha}}\right|_{x=x_{0}}\frac{1}{2\sigma}
\end{equation}
together with (\ref{pproj}) and (\ref{w}) proves that at any point $\varphi_{0} \in M^{\sigma}_{3,3}$, the spatial and momentum space components of $\frac{d\varphi}{dt}$ are related in the same way as their classical counterparts in the phase space. 
Furthermore, the derived relationships (\ref{pproj}), (\ref{w}), (\ref{spreadcomp}), (\ref{decomposition}) and (\ref{Va}) remain true at $t=0$ even when the potential $V$ depends on time. In fact, the only expression that contains time derivatives of $V$ is the derivative $\frac{d^{2}r}{dt^{2}}$ in (\ref{Va}). However, the corresponding terms $\pm \frac{i}{2r}\frac{dV}{dt}$ cancel out because of the reality of $\frac{d^{2}r}{dt^{2}}$.

The immediate consequence of these results and the linear nature of the Schr{\"o}dinger equation is that under the Schr{\"o}dinger evolution with the Hamiltonian ${\widehat h}=-\frac{\hbar^{2}}{2m}\Delta+V({\bf x},t)$, the state constrained to $M^{\sigma}_{3,3}$ moves like a point in the phase space representing a particle in Newtonian dynamics. That is, if at each $\varphi_{0} \in M^{\sigma}_{3,3}$, the components of $-\frac{i}{\hbar}{\widehat h}\varphi_{0}$ that are orthogonal to $M^{\sigma}_{3,3}$ are made to vanish while the tangent components are preserved, then the state $\varphi$ will move according to classical physics.  So,
{\it Newtonian dynamics of a particle is the dynamics of one-particle quantum system with state constrained to $M^{\sigma}_{3,3}$}.

On the other hand, there is a unique unitary evolution (one parameter group of unitary operators) on $L_{2}(\R^{3})$, which, being restricted to $M^{\sigma}_{3,3}$, under projections (\ref{pproj}), (\ref{w}) yields the Newtonian values of velocity and acceleration. In fact, equations (\ref{pproj}), (\ref{w}) for the states $\varphi$ given by (\ref{initial}) imply the Ehrenfest theorem
\begin{equation}
\label{E1}
2\mathrm{Re} \left(\frac{d\varphi}{dt}, {\widehat x} \varphi \right)=\left(\varphi, \frac{\widehat p}{m}\varphi \right)
\end{equation}
and
\begin{equation}
\label{E2}
2\mathrm{Re} \left(\frac{d\varphi}{dt}, {\widehat p} \varphi \right)=\left(\varphi, - \nabla V({\bf x}) \varphi \right).
\end{equation}
But the set  $M^{\sigma}_{3,3}$ of such vectors $\varphi$ is complete in $L_{2}(\R^{3})$ and on a complete set  the Ehrenfest theorem  (\ref{E1}) and (\ref{E2}) together with the condition of unitarity of evolution is known to imply the Schr{\"o}dinger equation. So formulas  (\ref{pproj}), (\ref{w}) on $M^{\sigma}_{3,3}$ imply the Schr{\"o}dinger dynamics of the state of the particle on the space of states.

The analogous results can be derived for systems of $n$-classical particles. For instance, consider a system of two distinguishable particles, described by the usual Hamiltonian 
\begin{equation}
\label{ham2}
{\widehat h}=-\frac{\hbar^2}{2m_1}\Delta_{1}-\frac{\hbar^2}{2m_2}\Delta_{2}+V({\bf x_{1}}, {\bf x_{2}}),
\end{equation}
where the indices $1$ and $2$ refer to the corresponding particles. 
The set of states $\varphi_{1}\otimes \varphi_{2}$, where $\varphi_{1}$ and $\varphi_{2}$ for each particle are of the form (\ref{initial}) is a $12$-dimensional embedded submanifold $M^{\sigma}_{6,6}$ of the Hilbert space $L_{2}(\R^{3})\otimes L_{2}(\R^{3})$ with induced Riemannian metric, isometric to the classical phase space $\R_6 \times \R_6$ of the two-particle system. Vectors 
\begin{equation}
\label{set1}
\left(-\frac{\partial r_{1}}{\partial x_{1}^{k}}e^{i\frac{{\bf p}_{1}({\bf x}_{1}-{\bf a}_{1})}{\hbar}}\right)\otimes \varphi_{2}, \quad i\frac{\partial \theta_{1}}{\partial p_{1}^{j}}\varphi_{1}\otimes \varphi_{2}
\end{equation}
and
\begin{equation}
\label{set2}
\varphi_{1}\otimes \left(-\frac{\partial r_{2}}{\partial x_{2}^{k}}\right)e^{i\frac{{\bf p}_{2}({\bf x}_{2}-{\bf a}_{2})}{\hbar}}, \quad \varphi_{1}\otimes i\frac{\partial \theta_{2}}{\partial p_{2}^{j}} \varphi_{2}
\end{equation}
are tangent to the phase spaces $M^{\sigma}_{3,3}\otimes \varphi_{2}$ and $\varphi_{1}\otimes M^{\sigma}_{3,3}$ of individual particles. These vectors are orthogonal for all values of $k,j=1,2,3$ and form a basis in the space tangent to $M^{\sigma}_{6,6}$.

Suppose now that a two particle quantum system has initial state in $M^{\sigma}_{6,6}$ and evolves by the Hamiltonian (\ref{ham2}). Because each operator $\Delta_{k}$ acts on just one function in the tensor product $\varphi_{1}\otimes \varphi_{2}$ and because the inner product in $L_{2}(\R^{3})\otimes L_{2}(\R^{3})$ is the product of inner products for individual particles, it follows that the components of the velocity vector $\frac{d}{dt} \left(\varphi_{1}\otimes \varphi_{2}\right)$ in the basis (\ref{set1}), (\ref{set2}) are given for each particle by their Newtonian values. For instance,
\begin{equation}
    \left(\frac{d\varphi_{1}}{dt} \otimes \varphi_{2}, -\frac{\partial r_{1}}{\partial x_{1}^{k}}e^{i\frac{{\bf p}_{1}({\bf x}_{1}-{\bf a}_{1})}{\hbar}}\otimes \varphi_{2}\right)=\frac{v_{1}^{k}}{2\sigma_{1}},
\end{equation}
where ${\bf v_{1}}={\bf p}_{1}/m_{1}$, etc.
It follows that: 
\newline
{\it Newtonian dynamics is the dynamics of a $n$-particle quantum system whose state is constrained to the phase-space submanifold $M^{\sigma}_{3n,3n}$ of the space $L_{2}(\R^{3})\otimes \ ... \ \otimes L_{2}(\R^{3})$ consisting of tensor product states $\varphi_{1}\otimes \ ... \ \otimes \varphi_{n}$ with $\varphi_{k}$ of the form} (\ref{initial}).

\section{Quantum probability and the classical normal distribution}

If a classical experiment for measuring the position of a macroscopic particle is performed, the result is generically a normal probability distribution of the position variable. Now the classical space $\R^{3}$ is identified with the submanifold $M^{\sigma}_{3}$ in the Hilbert space $L_{2}$ of states (equivalently, with the submanifold $M_{3}$ in the space ${\bf H}$). 
A macroscopic particle is identified with a quantum system constrained to the phase space $M^{\sigma}_{3,3}$.   
Measuring position of a macroscopic particle can be then described in terms of states in $S^{L_{2}}$. Because of this, the normal distribution of position of a macroscopic particle and the probability of transition between quantum states of a microscopic particle become related. It will be shown that, under measurements, macroscopic and microscopic particles obey the same law. Namely:

\smallskip

\noindent {\it The Born rule for a position measurement of a microscopic particle implies the normal probability distribution of position of a macroscopic particle.}

\smallskip

\noindent{\it Conversely, suppose that measurements of position of a macroscopic particle are distributed normally. Suppose further that the probability $P(\varphi, \psi)$ for a microscopic particle in an arbitrary state $\varphi\in L_{2}$  to be found under a measurement in a state $\psi$ depends only on the distance $\rho(\pi(\varphi), \pi(\psi))$ between the states, in the Fubini-Study metric on the projective space $CP^{L_{2}}$. Then $P(\varphi, \psi)=\cos^{2}\rho(\pi(\varphi), \pi(\psi))$.}
\smallskip

\noindent{To summarize:}

\smallskip

\noindent {\it The normal probability distribution of a position random variable for a particle in the classical space implies the Born rule for transitions between arbitrary quantum states of the particle and vice versa.}
\smallskip

\noindent To prove this, note that a macroscopic particle is described in the classical phase space $\R^{3}\times \R^{3}=M^{\sigma}_{3,3}$, and so its state at a given time is given by the function (see (\ref{initial})): 
\begin{equation}
\label{del}
\varphi_{\bf a}({\bf x})=
\left(\frac{1}{2\pi \sigma^{2}}\right)^{3/4}e^{-\frac{({\bf x}-{\bf a})^{2}}{4\sigma^{2}}}e^{i\frac{{\bf p}({\bf x}-{\bf a})}{\hbar}}
\end{equation}
Let ${\widetilde \delta^{3}_{\bf a}}({\bf x})$ be the modulus $|\varphi_{\bf a}|$ and let $\delta^{3}_{\bf a}$ denote the usual delta-function. By the Born rule, the probability density $f({\bf b})$ to find the particle at a point ${\bf b}$  is equal to
\begin{equation}
\label{Born1}
f({\bf b})=|\varphi_{\bf a}({\bf b})|^{2}=|({\widetilde \delta^{3}_{\bf a}}, \delta^{3}_{\bf b})|^{2}=\left(\frac{1}{2\pi \sigma^{2}}\right)^{3/2}e^{-\frac{({\bf a}-{\bf b})^{2}}{2\sigma^{2}}},
\end{equation}
which is the normal distribution function. It follows that on the elements of $M^{\sigma}_{3}$, the Born rule {\it is} the rule of normal distribution.

Conversely, assume the normal probability distribution of position measurements for macroscopic particles. Here it will be sufficient to deal with particles at rest.
A macroscopic particle at rest is represented by the state ${\widetilde \delta^{3}_{\bf a}}({\bf x})$ (zero phase) in the classical space $\R^{3}=M^{\sigma}_{3}$, which is a submanifold of $M^{\sigma}_{3,3}$. It was shown that the Born rule and the normal distribution law are the same for the states in $M^{\sigma}_{3,3}$, in particular, for the states ${\widetilde \delta^{3}_{\bf a}}({\bf x})$. Therefore, the normal distribution rule can be also written in the form of the Born rule
\begin{equation}
\label{Born2}
P(\tilde{\delta^{3}_{\bf a}}, \tilde{\delta^{3}_{\bf b}})=|(\tilde{\delta^{3}_{\bf a}}, \tilde{\delta^{3}_{\bf b}})|^{2},
\end{equation}
where $P(\tilde{\delta}^{3}_{\bf a}, \tilde{\delta}^{3}_{\bf b})$ is the probability of transition from the state $\tilde{\delta}^{3}_{\bf a}$ to the state $\tilde{\delta}^{3}_{\bf b}$ under a measurement of an appropriate observable.
Note that (\ref{Born1}) is the probability density while (\ref{Born2}) is the probability of transition. However, assuming ${\widetilde \delta^{3}}_{\bf b}$ is sufficiently sharp, the formulas mean the same thing. In fact, in this case $\delta^{3}_{\bf b}$ in (\ref{Born1}) can be replaced with ${\widetilde \delta^{3}}_{\bf b}$. For this recall that ${\widetilde \delta^{3}}_{\bf b}$ is unit-normalized in $L_{2}(\R^{3})$:
\begin{equation}
\int |{\widetilde \delta^{3}}_{\bf b}({\bf x})|^{2}d^{3}{\bf x}=1. 
\end{equation}
Let $h$ be the height ${\widetilde \delta^{3}}_{\bf b}({\bf b})$ of ${\widetilde \delta^{3}}_{\bf b}$ and let $\Delta x$ be defined by
\begin{equation}
h^{2}\cdot (\Delta x)^{3}=\int |{\widetilde \delta^{3}}_{\bf b}({\bf x})|^{2}d^{3}{\bf x}=1. 
\end{equation}

Then $h=\frac{1}{(\Delta x)^{3/2}}$ and
\begin{equation}
\label{Born3}
|({\widetilde \delta^{3}}_{\bf a}, \delta^{3}_{\bf b})|^{2} \approx \left|{\widetilde \delta^{3}}_{\bf a}({\bf b}) \int \frac{1}{(\Delta x)^{3/2}} d^{3}{\bf x} \right|^{2},
\end{equation}
where integration is over the cube of side $\Delta x$ centered at ${\bf b}$. As a result,
\begin{equation}
\label{Born4}
|({\widetilde \delta^{3}}_{\bf a}, {\widetilde \delta^{3}}_{\bf b})|^{2} \approx \left|{\widetilde \delta^{3}}_{\bf a}({\bf b}) \right|^{2}(\Delta x)^{3}=f({\bf b})(\Delta x)^{3},
\end{equation}
which relates the probability in (\ref{Born2}) to the normal probability density in (\ref{Born1}) and identifies $P(\tilde{\delta^{3}_{\bf a}}, \tilde{\delta^{3}_{\bf b}})$ with the probability of finding the macroscopic particle near the point ${\bf b}$. 

The Born rule (\ref{Born2}) can be also written as
\begin{equation}
\label{PP}
P(\tilde{\delta^{3}_{\bf a}}, \tilde{\delta^{3}_{\bf b}})=\cos^{2}\rho(\tilde{\delta^{3}_{\bf a}}, \tilde{\delta^{3}_{\bf b}}),
\end{equation}
where $\rho(\tilde{\delta^{3}_{\bf a}}, \tilde{\delta^{3}_{\bf b}})$ is the distance between the states $\tilde{\delta^{3}_{\bf a}}, \tilde{\delta^{3}_{\bf b}}$ in the Fubini-Study metric on the projective space $\pi: S^{L_{2}}\longrightarrow CP^{L_{2}}$. Here $\pi(\tilde{\delta^{3}_{\bf a}})$ is identified with $\tilde{\delta^{3}_{\bf a}}$, which is possible because the state is real-valued.

The Fubini-Study distance between the states $\tilde{\delta^{3}_{\bf a}}$, $\tilde{\delta^{3}_{\bf b}}$ takes on all values from $0$ to $\pi/2$, which is the largest possible distance between points in $CP^{L_{2}}$. By assumption, the probability $P(\varphi, \psi)$ of transition between any states $\varphi$ and $\psi$ depends only on the Fubini-Study distance 
$\rho(\pi(\varphi), \pi(\psi))$ between the states. Given arbitrary states 
$\varphi, \psi \in S^{L_{2}}$, let then $\tilde{\delta^{3}_{\bf a}}$,
$\tilde{\delta^{3}_{\bf b}}$ be two states in $M^{3}_{\sigma}$, such that
\begin{equation}
\rho(\pi(\varphi), \pi(\psi))=\rho(\tilde{\delta^{3}_{\bf a}}, 
\tilde{\delta^{3}_{\bf b}}).
\end{equation}
From the assumed normal probability distribution for the states $\tilde{\delta^{3}_{\bf a}}$ and the assumption that probability of transition depends only on the Fubini-Study distance between the states, it then follows that
\begin{equation}
P(\varphi, \psi)=P(\tilde{\delta^{3}_{\bf a}}, \tilde{\delta^{3}_{\bf b}})=\cos^{2}\rho(\tilde{\delta^{3}_{\bf a}}, \tilde{\delta^{3}_{\bf b}})
=\cos^{2}\rho(\pi(\varphi), \pi(\psi)),
\end{equation} 
which yields the Born rule for arbitrary states. This proves the claim.

This beautiful result is based on a highly non-trivial way in which the classical space is embedded into the Hilbert space of states. Namely, because of the special properties of the embedding, the "classical law" (normal distribution of observation results) becomes a part of the quantum law, which simply extends the classical law to superpositions. The extension is unique if the assumption is made that the probability of transition must only depend on the distance between states in the Fubini-Study metric. 

In more detail, denote the distance between two points ${\bf a}, {\bf b}$ in $\R^{3}$ by $\left\|{\bf a}-{\bf b}\right\|_{\R^{3}}$. Under the embedding of the classical space into the space of states, the variable ${\bf a}$ is represented by the state $\tilde{\delta}^{3}_{\bf a}$. The set of states $\tilde{\delta}^{3}_{\bf a}$ form a submanifold $M^{\sigma}_{3}$ in the Hilbert spaces of states $L_{2}(\R^{3})$. The manifold $M^{\sigma}_{3}$ is "twisted" in $L_{2}(\R^{3})$, it belongs to the sphere $S^{L_{2}}$ and spans all dimensions of $L_{2}(\R^{3})$. Distance between the states $\tilde{\delta}^{3}_{\bf a}$, $\tilde{\delta}^{3}_{\bf b}$ in $L_{2}(\R^{3})$ or in the projective space $CP^{L_{2}}$ is not equal to $\left\|{\bf a}-{\bf b}\right\|_{\R^{3}}$. In fact, the former distance measures length of a geodesic between the states while the latter is obtained using the same metric on the space of states, but applied along a geodesic in the twisted manifold $M^{\sigma}_{3}$. In precise terms the relation between the two distances is given by
\begin{equation}
\label{main}
e^{-\frac{({\bf a}-{\bf b})^{2}}{4\sigma^{2}}}=\cos^{2}\rho(\tilde{\delta}^{3}_{\bf a}, \tilde{\delta}^{3}_{\bf b}),
\end{equation}
where the left hand side is a result of integration in (\ref{Born2}). This equation is what accounts for the relation between the normal probability distribution and the Born rule.

\section{Summary}

The classical space and classical phase space are now embedded into the space of states of the corresponding quantum system and form a complete set (overcomplete basis) in that space. The dynamics of a classical $n$-particle mechanical system is identified with the Schr{\"o}dinger dynamics constrained to the classical phase space. Conversely, there is a unique unitary time evolution on the space of states of a quantum system that yields Newtonian dynamics when constrained to the classical phase space. The normal distribution law is derived from the Born rule. Conversely, the Born rule is the only probability law on the the projective space of states that is isotropic and yields the normal distribution on a classical configuration submanifold.
These results suggest that other areas of tension between classical and quantum physics can be now fruitfully explored.


\end{document}